\documentstyle[12pt,aaspp4,flushrt]{article}
\newcommand{\gW}{$\Omega$}
\newcommand{\gs}{$\sigma_{8}$}
\newcommand{\gm}{$\pm$}
\newcommand{\gi}{$\sim$}
\newcommand{\gie}{$\simeq$}
\newcommand{\gh}{$\rm h^{-1}$}
\newcommand{\gE}{{\em et al.}}

\begin{document}

\title{The Most Massive Distant Clusters: \\
Determining \gW\ and \gs}
\author{Neta A. Bahcall and Xiaohui Fan\\
Princeton University Observatory\\
Princeton, NJ 08544\\
electronic mail : neta,fan@astro.princeton.edu}

\section*{Abstract}
The existence of the three most massive clusters of galaxies observed so far at $z>0.5$ is used to
constrain the mass density parameter of the universe, \gW, and the amplitude
of mass fluctuations, \gs. 
We find \gW=$0.2_{-0.1}^{+0.3}$, and \gs = $1.2_{-0.4}^{+0.5}$ (95 \%).
We show that the existence of even the {\em single}  
most distant cluster at $z$=0.83,
MS1054--03, with its large gravitational lensing mass,
high temperature, and large velocity dispersion, 
is sufficient to establish powerful constraints. 
High-density, \gW =1 (\gs\gie0.5--0.6) Gaussian models are ruled out by these data ($\lesssim 10^{-6}$ probability); 
the \gW=1 models predict only $\sim 10^{-5}$ massive clusters at $z > 0.65$ 
($\sim 10^{-3}$  at $z > 0.5 $) instead of the 1 (3) clusters  observed.\\ \\
{\em subject headings :}cosmology : observation - - - cosmology : theory - - -
galaxies : cluster : general - - - galaxies : evolution - - -
large scale structure of universe

\section{Introduction}

The observed present-day abundance of rich clusters of galaxies places a
strong constraint on cosmology: \gs$\Omega^{0.5} \simeq 0.5$, where \gs\ is the {\em rms} mass
fluctuations on 8 \gh\ Mpc scale, and \gW\ is the present 
cosmological density parameter (Henry \& Arnaud 1991, Bahcall \& Cen 1992,
White \gE\ 1993, Eke \gE\ 1996, Viana \& Liddle 1996, Pen 1997, Kitayama
\& Suto 1997). This constraint is degenerate in \gW\ and \gs; models with \gW =1,
\gs \gi 0.5 are indistinguishable from models with \gW \gi  0.25, \gs \gi 1.
(A \gs \gie 1 universe is unbiased, 
with mass following light on large scales;
\gs \gie 0.5 implies a mass distribution  wider than light).

The {\em evolution} of cluster abundance with redshift, especially for massive
clusters, breaks the degeneracy between \gW\ and \gs\
(see, e.g., Peebles et al.  1989, Oukbir \& Blanchard 1992, 1997,  Eke \gE\ 1996, Viana \& Liddle
1996, Carlberg \gE\ 1997, Bahcall \gE\ 1997, Fan \gE\ 1997, 
Henry 1997). 
The evolution of high mass clusters is strong in \gW\ =1, low-\gs\ (biased)
Gaussian models, where only a very low cluster abundance is expected at $z>$0.5.
Conversely, the evolution rate in low-\gW\ , high-\gs\ models is
mild and the cluster abundance at $z>$0.5 is much higher than in \gW=1
models. 
In Bahcall \gE\ (1997) and Fan \gE\ (1997) we used the CNOC cluster sample (Carlberg \gE\ 1997a,b,
Luppino \& Gioia 1995) to $z \lesssim$ 0.5 -- 0.8 
(with measured masses to $z \lesssim 0.5$) to decouple \gW\ and \gs:
we found \gW \gie  0.3 \gm 0.1 and \gs \gie 0.83 \gm 0.15,
consistent with Carlberg \gE (1997a).
The evolution rate, and the
distinction among cosmological models, increases with cluster mass and with
redshift: in \gW\ =1, low-\gs\  models, 
very massive clusters are not expected 
to exist at high redshifts. 

In the present paper we extend the previous studies to larger mass and
higher redshift clusters, using the three most massive
clusters observed so far
at high redshifts ($z$ \gie 0.5--0.9) to independently constrain \gW\ and \gs. 
The clusters discussed in this paper are the three most massive distant
clusters from the EMSS/CNOC sample used above, with masses larger by a 
factor of $\sim$ 2 than the mass-threshold used 
previously (Evrard 1989, Bahcall \gE\ 1997, Fan \gE\ 1997, Carlberg \gE\ 1997a).
Reliably measured masses are now available for these clusters from 
gravitational lensing, temperatures, and velocity dispersions,
not previously available in the above studies.
Strong Sunyaev-Zel'dovich decrements have also been observed for these clusters,
further suggesting that these are massive clusters with large amount of gas.
The three clusters  have the highest
masses (from weak lensing observations), the highest velocity dispersions 
($\sigma_{r} \gtrsim $1200\ km $s^{-1}$), and the highest temperatures (T$\gtrsim $8 kev) 
in the $z >$ 0.5 EMSS survey (\S 2).
Therefore, they provide a strong constraint  on cosmology.
We discuss the cluster data in \S 2 and the cosmological implications in \S 3.
A Hubble constant of $\rm H_{0}=100\ h\ km\ s^{-1} Mpc^{-1}$  is used.

\section{Massive Distant Clusters}

We use the three most massive luminous clusters from the six $z > 0.5$ X-ray
clusters in the EMSS Survey (Henry {\em et al.} 1992,
Luppio \& Gioia 1995). Extensive data are now available for these three
clusters : temperatures (T $\gtrsim 8$ kev), 
velocity dispersions ($\rm \sigma_{r} \gtrsim 1200\ km\ s^{-1}$),
 and masses ($\rm M(\leq 0.5 h^{-1} Mpc) \gtrsim 5 \times 10^{14} h^{-1} M_{\odot}$,
from weak gravitational lensing observed for two of the
three clusters). 
A summary of the data is presented in Table 1.

 The mass of the most distant cluster, MS1054--03 at $ z=0.83 $, has been
determined from weak lensing (Luppino \& Kaiser 1997): 
$\rm M_{GL} (\leq 0.82 h^{-1} Mpc) \simeq 10^{15} h^{-1} M_{\odot}$ 
(within $\rm R = 0.82 h^{-1} Mpc = 1.5 h^{-1}$ Mpc-{\em comoving}-radius).
This mass assumes the 
 background galaxies are located at the most realistic redshift range
of $z \sim 1.5 - 3 $ (Luppino \& Kaiser 1997),
and takes account of the surface mass zero point (Luppino \& Kaiser 1997) 
and the de-projection to  spherical mass within R 
(as given by Hjorth {\em et al.} 1997).
The uncertainty in this mass estimate is $\pm 30 \%$. 
(If the background galaxies are located at a mean redshift of $z \sim 1.5$,
the cluster mass is even larger, by 35 \%; Luppino \& Kaiser 1997.)
The evolution analysis presented in this paper requires only a best-estimate of 
a {\em THRESHOLD MASS} for the clusters (within a given radius); 
i.e., cluster masses need only be $\rm \geq M(threshold) \pm \Delta M (threshold)$.
As long as the clusters are above the given mass threshold, 
accurate knowledge of the cluster mass is unimportant.
We determine a conservative best-estimate mass threshold later on.

MS1054--03 is at least twice as massive as the Coma cluster (at the same radius). 
The independent mass estimates from the cluster temperature (12.3 kev) 
and velocity dispersion ($\rm \sigma_{r} \simeq 1360\  km\  s^{-1}$,
Donahue \gE\ 1997), assuming hydrostatic equilibrium, agree well with the
lensing mass (Table 1). These estimates use the observed mean M(T) and M($\sigma_{r}$)
mass relations (see e.g., Bahcall \& Fan 1998; also Hjorth \gE\ 1997, 
Carlberg \gE\ 1997a,b): 
$\rm M(\leq1 h^{-1} Mpc) \simeq T_{kev} \times 10^{14} h^{-1} M_{\odot}$ and 
$\rm (M\leq 1.5 h^{-1} Mpc) \simeq 2.4\sigma_{r}^{2}\times 1.5 h^{-1} (Mpc)/G$.
Mass uncertainties are estimated from the observed uncertainties in T and $\sigma_{r}$ and
the $\sim 20 \%$ {\em rms} scatter in the observed relations.
Cluster masses can be extended to other radii R 
using the observed cluster profile: 
$\rm M(\leq R) \propto R$ for $\rm R \sim 0.5 - 1 h^{-1} Mpc$ and 
$\rm M(\leq R) \propto R^{0.64}$ for $\rm R \sim 1-2 h^{-1} Mpc$
(Carlberg \gE\ 1997b; also Fischer \& Tyson 1996).
The results are only weakly dependent on the exact profile slope;
changing the slope by $\pm 0.3$ changes the results by $\lesssim 15 \%$.

 The other two clusters, MS0016+16 and MS0451--03,
are the two most luminous X-ray clusters in the $z>0.5$ EMSS sample; they
have high velocity dispersions ($\rm \sigma_{r} \gtrsim 1200\ km\ s^{-1}$) and high temperatures
($\gtrsim$ 8 kev). The mass of MS0016+16 has been determined from weak lensing
(for R $\lesssim  0.3 \rm h^{-1}$ Mpc; Smail \gE\ 1995), and is consistent (when
extrapolated to the same radius) with the mass obtained from the cluster
velocity dispersion and temperature (Table 1). Strong
Sunyaev-Zel'dovich decrements have also been observed for all three clusters (Carlstrom 1997). 
The fact that the independent mass determinations (from lensing, 
temperature, and velocity dispersion) for the three clusters all
yield cluster masses that are consistent with each other (Table 1)
clearly indicates the reliability of these mass estimates.
The strong Sunyaev-Zel'dovich decrements detected in these clusters
further support this conclusion, indicating clusters with large amounts of gas.
The existence of substructure in any of these clusters will
not significantly affect the results, as shown by the agreement of all
three independent mass methods. Furthermore,  direct simulations  
show a $\lesssim 15 \%$ effect on the mass
determination due to substructure in comparable overdensity clusters (Evrard {\em et al.} 1996). 

In the following cosmological analysis, only a {\em mass threshold}
is needed for the clusters; 
the exact cluster mass above the threshold is unimportant.
Based on the data presented in Table 1, the best estimated mean cluster masses are 
10 \gm 2, 8.7 \gm 1.3, and $\rm 11 \pm 1.4 \times 10^{14} h^{-1} M_{\odot}$  for the three clusters (within $\rm 1.5 h^{-1} comoving\ Mpc$).
The uncertainties represent the formal errors for the weighted average cluster
mass as determined from the three independent methods: lensing, temperature,
and velocity dispersion.
We adopt a conservative mass 
threshold of $\rm M_{1.5-com} \simeq 8^{+2}_{-1} \times 10^{14} h^{-1} M_{\odot}$
(within $\rm 1.5 h^{-1} Mpc$ {\em comoving} radius). 
This threshold is safely near the low end of the observed cluster masses;
the mass threshold uncertainties are  included in the final results.
If the cluster masses are higher, the derived constraints become
even stronger.
Furthermore, even if one of the clusters is eliminated from the sample,
(e.g., if one, or even two cluster masses are below the threshold),
the main conclusions remain essentially unchanged.

     The number density of massive ($\geq 8 \times 10^{14} \rm h^{-1} M_{\odot}$)
 clusters  at $ z > 0.5$ is determined using the $\Sigma (1/V_{max})$ method 
 (e.g., Luppino \& Gioia 1995, Donahue \gE\ 1997). 
We find $n(z=0.5-0.65) \simeq 3^{+2.4}_{-1.9} \times 10^{-8} \rm h^{3} Mpc^{-3}$ (2 clusters)
and $n(0.65-0.9) \simeq 3.4^{+3.4}_{-2.8} \times 10^{-8} \rm h^{3} Mpc^{-3}$ (1 cluster) for \gW\ =1.
The error bars represent 68 \% confidence level assuming Poisson
statistics and equal likelihood for each $\log n$.
For \gW \gie 0.2, $n \simeq 2 \times 10^{-8} \rm h^{3} Mpc^{-3}$ for each of the two bins. (The density
may be larger by a factor of $\lesssim$ 2 if any of the other
$z>$0.5 EMSS clusters are as massive as these). 
How do these densities
compare with the abundance of $z \simeq 0 $ clusters? 
The abundance of $\rm M_{1.5} \geq 8 \times 10^{14} h^{-1} M_{\odot}$
clusters at $z \simeq$ 0 is taken from several sources: the mass
function of Bahcall \& Cen (1993); the temperature function of Henry
(1997) (see also Eke \gE\  1996, Viana \& Liddle 1996 and Pen 1997); and
the temperature function of Edge \gE\ (1990). The temperature
threshold corresponding to the above mass 
(within  $1.5 \rm h^{-1} comoving $ Mpc, at $z \simeq 0.05$) is T $\gtrsim$ 6.4 kev (see M(T,R) relation above). 
The  local cluster abundances are all in the range $\rm \sim 1.3 - 2 \times 10^{-7}
h^{3} Mpc^{-3} (\times 10^{\pm 0.3})$.
The velocity function of the ENACS survey (Fadda \gE\ 1996, Mazure \gE\ 1996,
 Borgani \gE\ 1997), 
at the relevant $\rm M_{1.5}$ threshold (corresponding to $\rm \sigma_{r} \gtrsim
1010\ km\ s^{-1}$) is
 $\sim 3 \times 10^{-7} \rm h^{3} Mpc^{-3}$
(based on 2 clusters in the complete $z \leq 0.08$ sample).

     The abundance of high mass clusters evolves
slowly from $z\sim $ 0.8 to $z \sim $ 0; the abundance at $z\sim 0.8$ is only a factor
of \gi\ 5--10 times lower than at present. (The evolution may in fact be
even slower since we conservatively included, as a lower limit,
  only the three best
studied most luminous clusters).
As we show below, the expected evolution rate in \gW\ =1 (low \gs) models
is $\sim 10^{5}$ times faster.

\section{Constraining \gW\ and \gs}

We compare the data with the expected evolution of cluster abundance
using the Press-Schechter (1974; P--S) formalism, which  describes the
evolution of the abundance of bound objects 
that grow from random-phase Gaussian initial fluctuations. The
P-S method yields results that are in good agreement
with simulations (down to the  simulation limit of
$\rm \sim 10^{-8} h^{3} Mpc^{-3};
$e.g., Eke \gE\  1996, Fan \gE\ 1997, Pen 1997).
P-S describes the evolution of the cluster mass function,

\begin{equation}
\frac{dn}{dM} = \left(\frac{2}{\pi}\right)^{\frac{1}{2}} \frac{\overline{\rho}}{
M^{2}} \frac{\delta_{c}}{\sigma(z,M)}\left|\frac{d\ln\sigma(z,M)}{d\ln M}\right|
\exp\left(\frac{-\delta_{c}^{2}}{2\sigma(z,M)^{2}}\right),
\end{equation}
     where $\sigma(z,M)$ is the linear theory {\em rms} mass density fluctuation in
spheres of mass M at redshift $z$, $\delta_{c} \simeq 1.68$ is the critical density
contrast needed for collapse (weakly dependent on \gW , Eke
\gE\ 1996, Kitayama \& Suto 1996), and $\overline{\rho}$ is the mean cosmic
density. The mass refers to the {\em virial mass} of the system.
The present {\em rms} mass fluctuation
within a sphere of mass M, $\sigma_{0}(M)$, relates to \gs\
as $\sigma_{0}(M) = \sigma_{8} ({M}/{M_{8}})^{-\alpha} \sim
\hspace{3mm} \sigma_{8}M^{-\alpha}\Omega^{\alpha}$, 
where $\alpha$ = (n+3)/6, n is the slope of the power spectrum at $\rm \sim 8 
h^{-1} Mpc$, and $\rm M_{8} \propto \Omega$
is the mean mass within a sphere of radius $8 \rm h^{-1} Mpc$.
We use $\rm n \simeq -1.4$, as observed
(corresponding to  $\rm \Omega h \simeq 0.25$
for a CDM spectrum); 
the results are insensitive to n and $\alpha$ (changing $\Omega$ by $< 20\%$
for reasonable changes in n; Fan \gE\ 1997).

We use three different (but not independent) approaches.
First, we calculate the evolution of the abundance of 
$\rm M_{1.5-com} \geq 8 \times 10^{14} h^{-1} M_{\odot}$
 clusters by integrating eq. (1).
We use the \gW--\gs\ constraint placed by the present-day cluster
abundance: \gs $\Omega^{0.45} = 0.53 \pm 0.05$ (for $\Lambda =0$) and 
\gs $\Omega^{0.53} = 0.53 \pm 0.05$ (for $\Omega + \Lambda =1$) 
 (Pen 1997; similar results by Eke \gE\ 1996; see also
Bahcall \& Cen 1992, White \gE\ 1993, Viana \& Liddle 1996, Kitayama \& Suto 1997); this
ensures that the proper normalization \gs\ is used for any \gW. The
virial mass in eq. (1) is converted to the observed $\rm M_{1.5-com}$ mass as
follows (see Fan \gE\ 1997): the virial overdensity is numerically
calculated for each \gW\ (\gs) model (e.g. Eke \gE\ 1996, Oukbir \&
Blanchard 1997), thus yielding the virial radius for a given virial
mass. The virial mass (i.e., the mass within the virial radius) is then
scaled to the 1.5 $\rm h^{-1}$ Mpc comoving radius using the observed
$\rm M(\leq R) \propto R^{0.64}$
profile (see \S 2). The results are insensitive to this
transformation;
changing the profile slope by $\pm 0.3$ changes the constraints on \gW\  by $\lesssim 15 \%$.

Figure 1 shows the  expected evolution of the cluster abundance as a function of
redshift  for different \gW, \gs\ combinations
that satisfy the present day cluster abundance. The
model curves range from \gW\ = 0.1 (\gs\ \gie\ 1.7) at the top of the figure (flattest, nearly
no evolution) to \gW\ = 1 (\gs\ \gie\ 0.5) at the bottom (steepest, strongest
evolution). The difference between low and high \gW\ models (i.e., 
high and low \gs) is dramatic for these high mass
clusters; the cluster abundance in \gW\ = 1 models  is 
$\sim 10^{-5}$ of the abundance in low \gW\ models. The data
exhibit {\em only a slow, relatively flat evolution}; this is expected {\em only
in low-\gW\ models}. The observed clusters, even just the single
 cluster at $z$ \gie\ 0.83, represent $ \sim 10^{5}$ more clusters than expected in any \gW\
=1 (\gs\ \gi\ 0.5--0.6) models such as Cold or
Mixed Dark Matter (CDM or MDM). The results
are similar for $\Lambda$ =0 and $\Lambda$ +\gW\  =1 models 
(the latter show slightly stronger evolution;
see Bahcall \gE\ 1997 and Fig. 2).

Figure  2 shows the allowed range (68 \% and 95 \%)
of \gW\ permitted by the existence of the massive clusters at $z=0.5 - 0.9$
(using the \gs--\gW\  normalization above). The expected
number of clusters with 
$\rm M_{1.5-com} \geq 8 \times 10^{14} h^{-1} M_{\odot}$
in the two redshift bins
in the
current sample is presented as a function of \gW\ (or \gs), and compared
with the two  and one detected clusters, 
respectively. 
The existence of the two 
clusters at $z=0.5 - 0.65$ is only possible if \gW\ \gie\ $0.25^{+0.15}_{-0.1}$,
\gs\ \gie\ 1.2 \gm\ 0.3 (95\%).
As \gW\ increases (\gs\ decreases), the expected number
of clusters drops exponentially. For \gW\ = 1 (\gs\ \gie\ 0.5), only $\sim 10^{-3}$
clusters are expected compared with the 2 observed,
representing a probability of $\sim 10^{-6}$. Similar
(but independent) results are obtained for the one cluster observed in the
$z$ = 0.65--0.9 bin (Fig. 2); the probability of finding one such cluster if
\gW\ = 1 is $\sim 10^{-5}$. These independent results reinforce
the above conclusions, 
ruling out \gW=1 models as very low probability  ($\lesssim 10^{-6}$).
This low probability ($\lesssim 10^{-6}$) applies for any  
mass-threshold $\rm M \gtrsim 7 \times 10^{14} h^{-1} M_{\odot}$;
the larger the mass, the lower the probability.
The difference between $\Lambda$ = 0 and $\Lambda$ +\gW\  = 1 models 
is small (Fig. 2).

A second method of analyzing the results, independent of the \gs\--\gW\
normalization, was discussed by Fan \gE\ (1997). We
showed that the evolution rate of cluster abundance depends
exponentially on \gs; if we ignore the normalization $\sigma_{8}\Omega^{0.5} \sim 0.5$  that
is needed to match the "absolute" cluster abundance at $z \sim 0$  and study
only the {\em rate} of the evolution profile with $z$ (i.e., its ``flatness''), we
can estimate \gs\ directly, nearly independent of \gW\ (as well as
nearly independent of the exact shape of the power spectrum and $\rm H_{0}$). 
Figure 3 shows the   
cluster abundance ratio , $n(z\simeq 0) / n(z\simeq 0.8)$, as a function of \gs\ (for 
 all \gW\ 's), determined from P-S (eq. 1). The very strong
evolution for low --\gs\ is clearly distinguished from the nearly no
evolution expected for \gs\ \gie\ 1 models (for any \gW\ ). 
The slow observed evolution rate (Fig.3)
requires a high \gs\ value : $\sigma_{8} \gtrsim 0.9$,
independent of \gW\ .
When combined with the proper normalization of the
present-day cluster abundance, a low \gW\ value is  obtained: $\Omega \lesssim 0.35$.
Similar results are found for $\Lambda=0$.

   A  third method of analyzing the results is to  determine an independent
\gW\ --\gs\ relation at high redshift (using P-S) based entirely
on the three high redshift clusters; this method is
independent of the $z$=0 cluster abundance. 
Figure 4 shows the allowed  \gW\ --\gs\ range determined by  
the high--$z$ cluster abundance.
The two  high redshift bins (shown by the solid and dashed curves in Fig. 4) 
are consistent  with each other.  They satisfy 
\begin{equation}
 \sigma_{8}\Omega^{0.24} \simeq 0.8 \pm 0.1 \hspace{3mm}(\Lambda = 0) 
,\hspace{3mm} \sigma_{8}\Omega^{0.29} \simeq 0.8 \pm 0.1
 \hspace{3mm} (\Lambda + \Omega =1) \hspace{3mm} [\rm for \hspace{2mm}z\simeq 0.5 - 0.9]
\end{equation}
     The dependence on \gW\ is weak.  These relations 
imply a {\em high \gs\ normalization}, \gs\ $\gtrsim$  0.8, for {\em any \gW\ $\leq 1$}.
The independent constraint provided by the present-day cluster
abundance (\S 3) is also presented in Fig. 4. These two \gW\ --\gs\
relations overlap only for low-\gW\ , high-\gs\ values. 
The allowed parameter ranges (68 \% and 95 \% confidence level contours) are shown in Fig 4b; 
the probability distributions are derived from the likelihood of each $\Omega-\sigma_{8}$
model assuming Poisson statistics.
We find \gW\ = $0.22^{+0.13}_{-0.07}$ and \gs=$1.18^{+0.24}_{-0.22}$ (for $\Lambda + \Omega =1$),
 and \gW=$0.17^{+0.14}_{-0.09}$ , \gs=$1.17^{+0.45}_{-0.25}$ (for $\Lambda = 0$)
(68 \% limits).
The 95\% limits are $\Omega = 0.22_{-0.1}^{+0.25}$, $\sigma_{8}=1.18^{+0.4}_{-0.32}$
($\Omega + \Lambda =1$) and
$\Omega = 0.17_{-0.1}^{+0.28}$, $\sigma_{8}=1.17^{+0.52}_{-0.32}$
($\Lambda=0$).
The results are consistent (at $\sim$ 1 $\sigma$ level) with those obtained
previously by Bahcall et al. (1997) and Fan et al. (1997) using lower
mass clusters,
and with the low-\gW\ conclusion of Luppino \& Kaiser (1997), Henry (1997),  and
Donahue \gE (1997).

In summary,  we find \gW\ =
     $0.2^{+0.3}_{-0.1}$ and \gs\ = $1.2_{-0.4}^{+0.5}$ (95\% C.L.). 
The high \gs\ value suggests a nearly unbiased universe, 
where mass approximately follows light on large scales.
The error-bars include an estimated uncertainty
in the cluster mass threshold ranging from
$\rm M_{1.5-com} \simeq 7$ to $\rm 10 \times 10^{14} h^{-1} M_{\odot}$.
The existence of even the single massive cluster
at $z$=0.83  places similar stringent constraints.
Using a lower mass threshold of $\rm M_{1.5-com} = 7  \times 10^{14} h^{-1} M_{\odot}$
increases $\Omega$ by 30 \%.
The results are insensitive to the power spectrum and to $\rm H_{0}$;
fixing h = $0.65 \pm 0.15$ (corresponding to $\rm n \simeq -1$ for
$\Omega = 1$, instead of $\rm n \simeq -1.4$)
lowers $\sigma_{8}$ by 10\% and increases \gW\ by 20\%.
To be consistent with \gW\ =1 models (such
as SCDM or MDM, \gs\ \gie\  0.5--0.6) {\em all three} cluster masses need to be unacceptably low,
$\rm M_{1.5-com} \sim 2-3 \times 10^{14} h^{-1} Mpc$ 
(i.e.  comparable to poor clusters of T \gi\ 2 -- 3 kev,
instead of the observed T $\sim 8 - 12$ kev);
this is excluded by the lensing, velocity dispersion,
temperature, and Sunyaev-Zel'dovich data.

We thank J. Hughes, G. Luppino, J. P. Ostriker, J. P. E. Peebles, D. N. Spergel 
and M. A. Strauss for helpful discussions. 
The work is supported by NSF grant AST93-15368.

\newpage

Table 1. Cluster Properties

\begin{footnotesize}
\begin{center}
\begin{tabular}{cccccccccc}\\ \hline
Cluster &z & $N_{0.5}^{(a)}$ & $T_{kev}^{(b)}$ & $\sigma_{r}^{(c)}$ & $\rm R_{1.5-com}$ & \multicolumn{4}{c}{$\rm M^{(d)}(\leq R_{1.5-com}) (10^{14} h^{-1} M_{\odot})$ } \\ 
 & & & &  ($\rm km\ s^{-1}$) & ($\rm h^{-1} Mpc)$ & $\rm M_{G.L.}$ & M(T) & M($\sigma$) & $\rm <M>$ \\ \hline
MS1054--03 & 0.83 & 50 \gm\ 10 & $12.3^{+3.1}_{-2.1}$  & 1360 \gm\ 450 & 0.82 & $\gtrsim 10 \pm 3$  & $10  \pm 3$ & $10 \pm 6$ & 10 \gm 2 \\ 
MS0016+16 & 0.55 & 66 \gm\ 16 & 8 \gm\ 1  & 1234 \gm\ 128  & 0.97 & $9 \pm 4 $ & $8  \pm 2 $ & $9 \pm 2$ & 8.7 \gm 1.3\\
MS0451--03 & 0.54 & 47 \gm\ 5  & $10.4 \pm 1.2$ & 1371 \gm\ 105 & 0.97 & -- & $10 \pm 2 $& $ 12\pm 2 $ & 11 \gm 1.4\\  \hline \hline
\end{tabular}
\end{center}
a) Central galaxy density count (Luppino \& Gioia 1995). ($N_{0.5}(Coma)=28$; Bahcall 1977)\\
b) Cluster temperature (Mushotzky \& Scharf 1997, Donahue 1996, Donahue \gE\ 1997; 90 \% error bars)\\
c) Cluster velocity dispersion (Carlberg \gE\ 1997b for $z \sim 0.5$ clusters, 68 \% error bars; Donahue \gE\ 1997 for $z \sim 0.8$ cluster,  90 \% error bar)\\
d) Cluster mass (within $\rm R_{com} = 1.5 h^{-1} Mpc$, column 6) as
determined from gravitational lensing ($\rm M_{G.L.}$; Smail \gE\ 1995, 
Luppino \& Kaiser 1997, Hjorth \gE\ 1997),
and from the cluster temperature and velocity dispersion of column 4--5
(M(T), M($\sigma$); \S 2). 
The weighted mean estimated mass for each cluster based on these data, $\rm <M>$, is also listed.
Only a {\em threshold} mass is needed for the evolution analysis; 
we adopt, conservatively, $\rm M_{1.5-com} \geq 8^{+2}_{-1} \times 10^{14} h^{-1} M_{\odot}$ for these clusters (\S 2).
\end{footnotesize}
\newpage
\begin{figure}
\vspace{-6cm}

\epsfysize=600pt \epsfbox{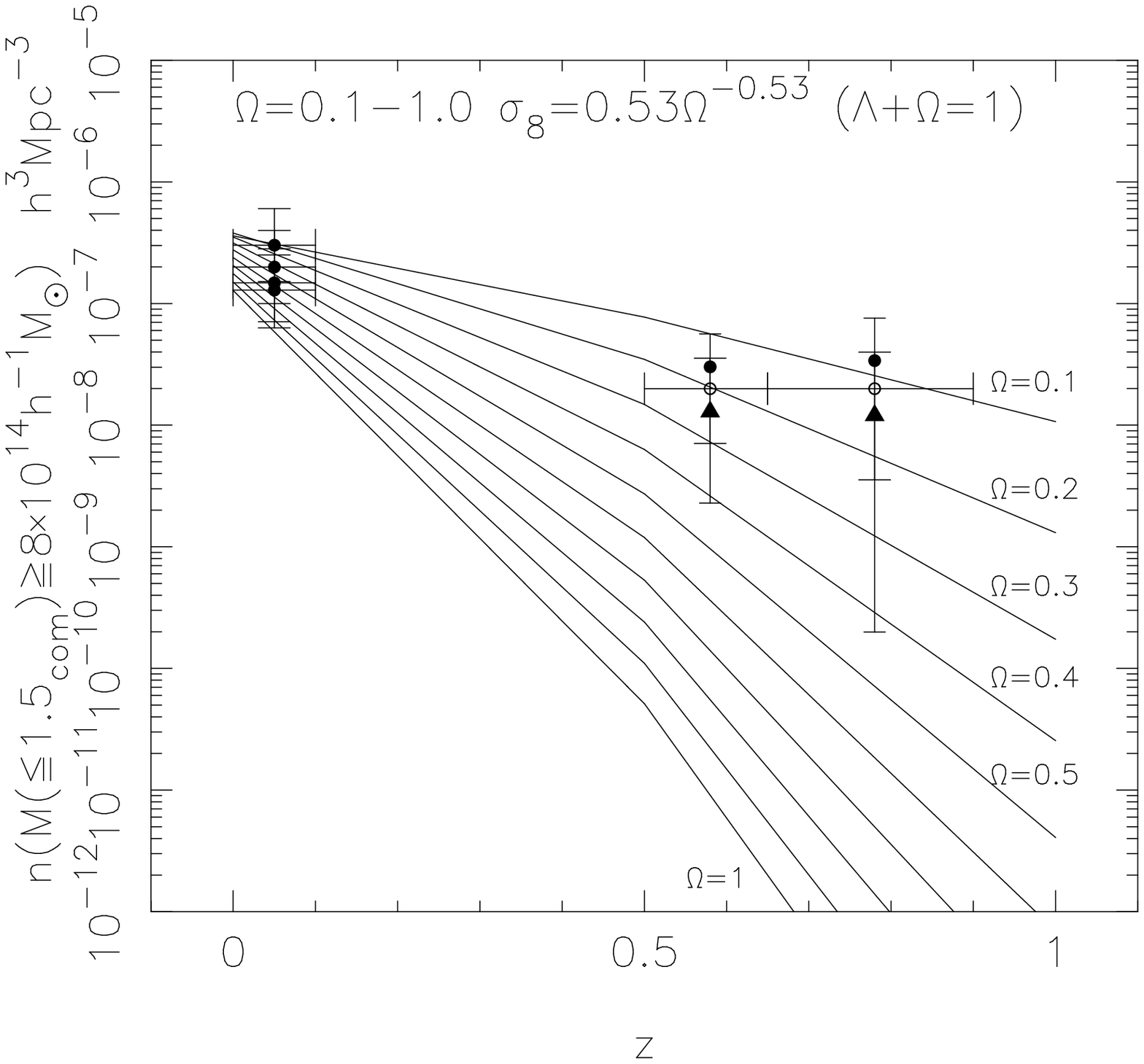}

\vspace{1cm}
Figure 1. Observed vs. model evolution of cluster abundance for $\rm M(\leq R_{com}=1.5 h^{-1} Mpc) \geq 8 \times 10^{14} h^{-1} M_{\odot}$ clusters.
The P-S model expectations (curves) are shown as a function of \gW(\gs) 
(for $\Omega + \Lambda = 1$; $\Lambda=0$ models yield similar results).
The $z \sim 0.6$ and 0.8 data points represent 2 and 1 clusters, respectively
(the most massive EMSS clusters; \S2);
filled circles are for \gW=1, open circles $\Omega \simeq 0.2$ ($\Lambda=0)$,
triangles $\Omega \simeq 0.2$ ($\Lambda = 0.8$).
The 68\% and 95\% error bars are shown.
(For $z \sim 0$, see \S 2).
\end{figure}
\begin{figure}
\vspace{-6cm}

\epsfysize=600pt \epsfbox{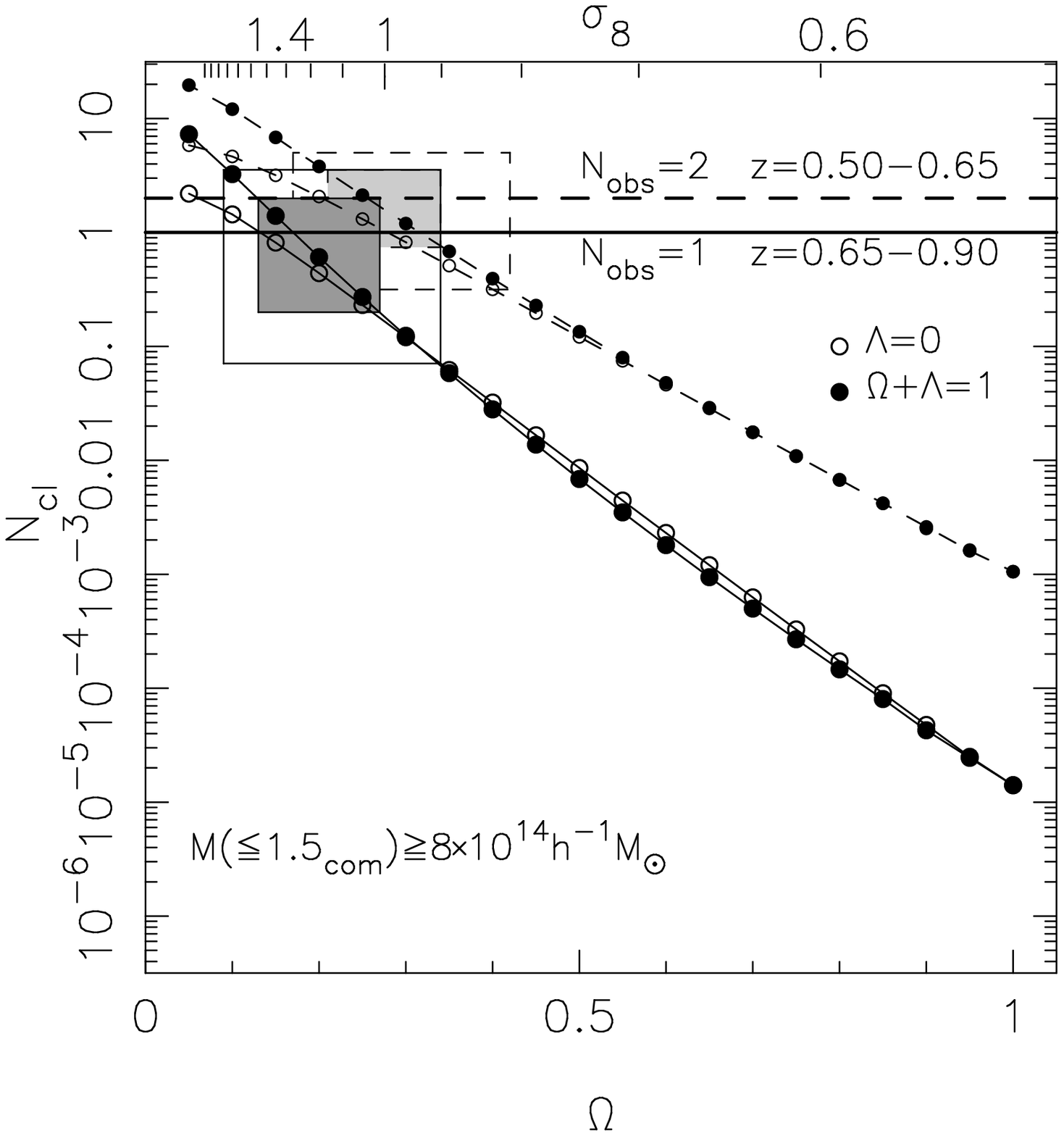}

\vspace{1cm}
Figure 2. Observed vs. expected (P-S) number of clusters in the survey in 
each of the two $z > 0.5$ bins (Fig 1), presented as a function of \gW\
(or \gs\ ).
Solid lines represent the $z=0.65 - 0.90$ bin (1 observed cluster);
dashed lines represent the $z=0.5 - 0.65$ bin (2 observed clusters).
The 68 \% and 95 \% confidence ranges are
marked by the dark and open regions near $\Omega \simeq 0.2, \sigma_{8} \simeq 1$.
The probability of $\Omega = 1$ models being consistent with either data bin  is $\sim 10^{-5}$.
\end{figure}
\begin{figure}
\vspace{-6cm}

\epsfysize=600pt \epsfbox{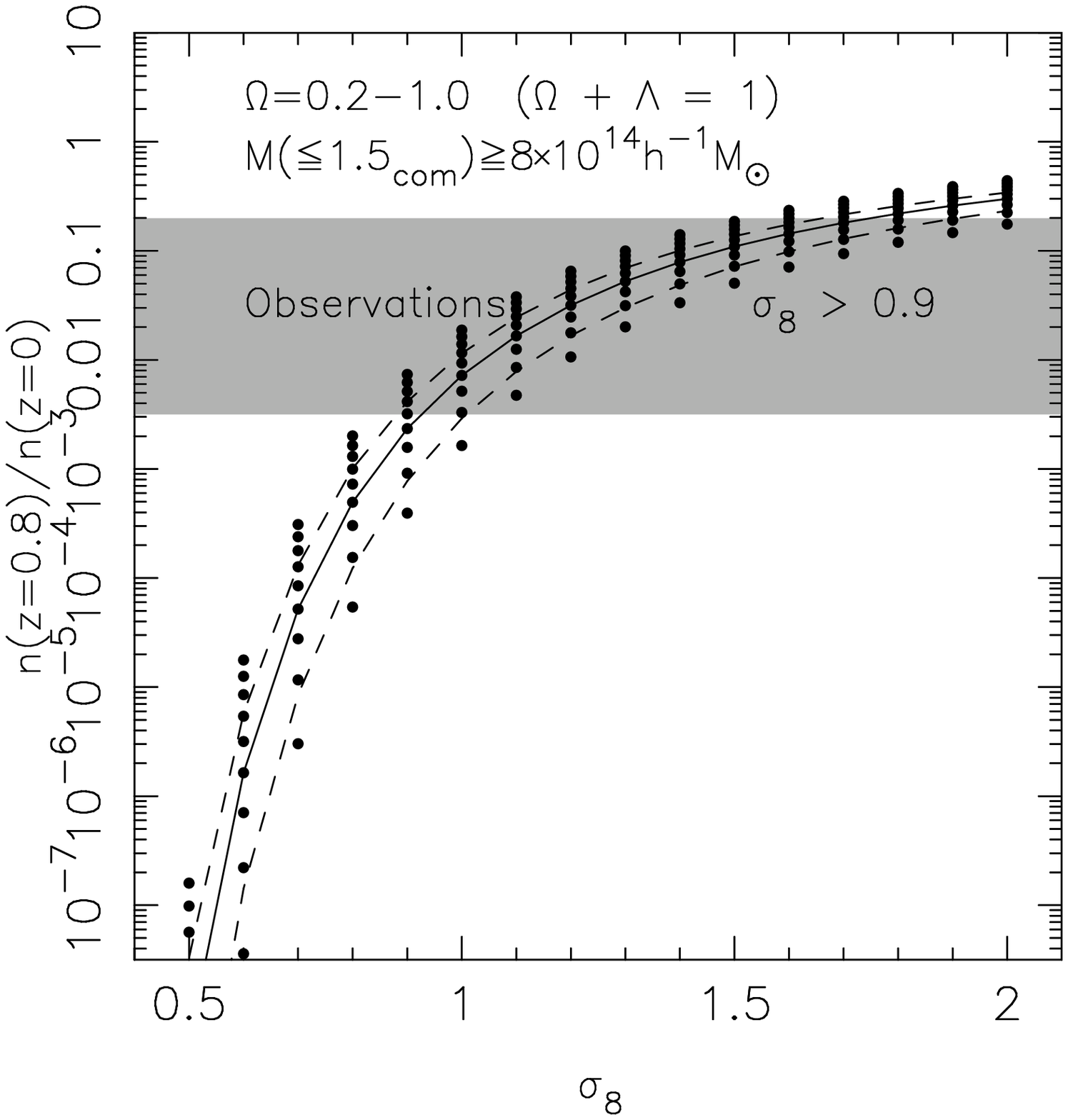}

\vspace{1cm}
Figure 3. Cluster abundance ratio, $n(z=0.8)/n(z=0)$,
vs. \gs\ from Press-Schechter (solid curve, 
for mean of all \gW's)  for
$\rm M_{1.5-com} \geq 8 \times 10^{14} M_{\odot}$ clusters.
Filled circles represent $\Omega$'s from 0.2 to 1 (bottom to top).
(Dashed curves represent the mass threshold range of 7 to 10 $\rm \times 10^{14} M_{\odot}$,
top and bottom, respectively).
The data (Fig 1, \S 2) are shown by the shaded region (68 \% level).
Similar results are obtained for $\Lambda=0$ (with weaker dependence on
$\Omega$).
\end{figure}
\begin{figure}
\vspace{-2.6cm}

\epsfysize=600pt \epsfbox{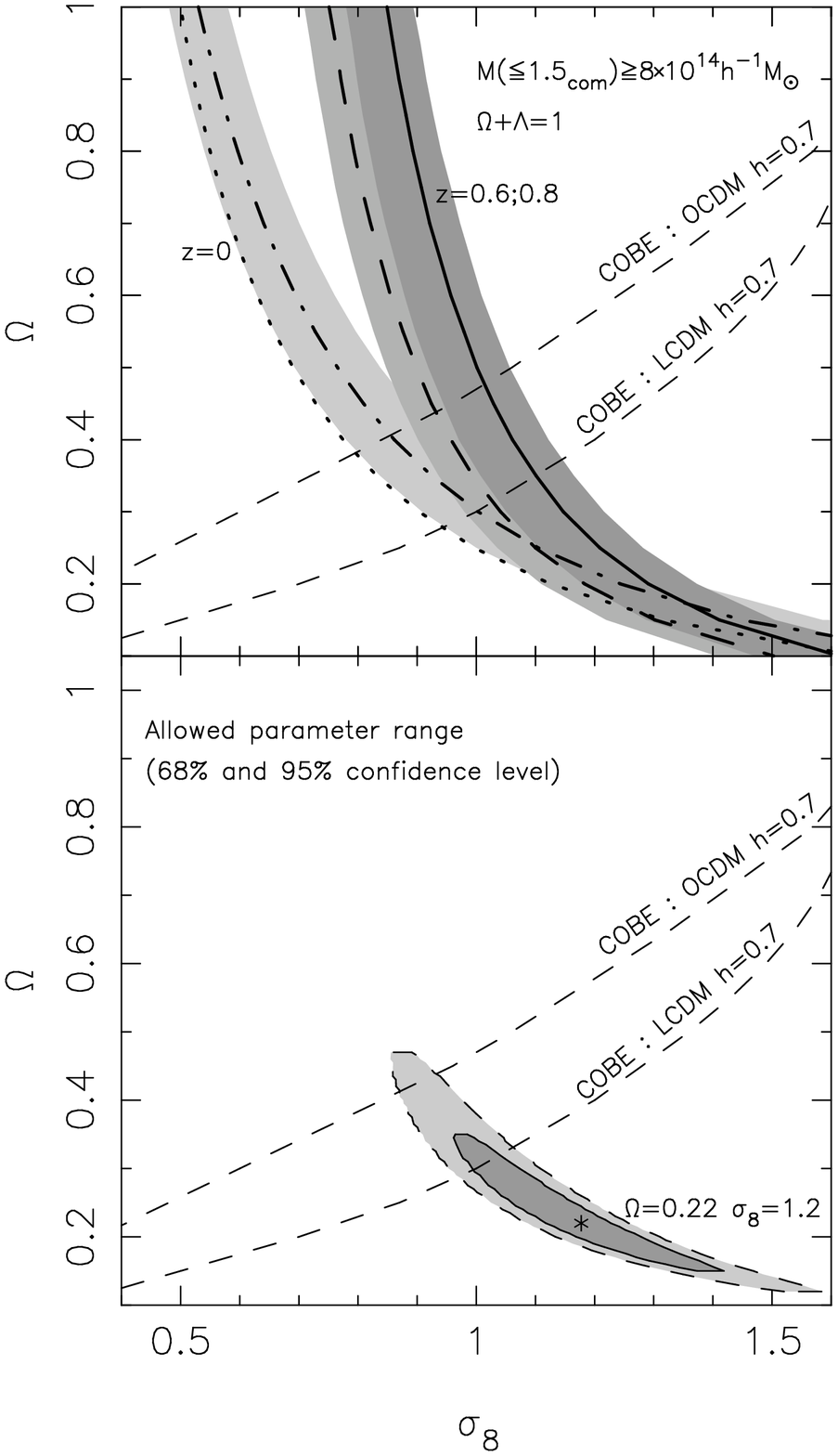}

Figure 4. Constrainting  \gW\ and \gs. 
The $z=0$ band represents the present-day cluster abundance constraint
($\sigma_{8} \Omega^{0.5} \simeq 0.5$; Pen 1997 dot-dashed curve,
Eke \gE\ 1996 dotted curve).
The $z=0.6, 0.8$ bands (dashed and solid curves) are the
constraints provided by the present high redshift clusters 
($\sigma_{8} \Omega^{0.29} \simeq 0.8$; eq 2.) (68\%).
The allowed parameter ranges (68 \% and 95 \%)
are shown (bottom figure).
Similar results are obtained for $\Lambda=0$ (\S 3).
The COBE normalization (Bunn \& White, 1997) are also shown 
(open and $\Lambda$ models).
\end{figure}

\newpage
\begin{thebibliography}{}

\bibitem{}Bahcall, N.A., 1977, ARA\&A, 15, 502

\bibitem{}Bahcall, N.A., \& Cen, R., 1992, ApJ, 398, L81

\bibitem{}Bahcall, N.A., \& Cen, R., 1993, ApJ, 409, L48

\bibitem{}Bahcall, N.A., Fan, X. \& Cen, R., 1997, ApJ, 485, L53

\bibitem{}Bahcall, N.A., \& Fan, X., 1998, in prep.

\bibitem{}Borgani, S., Gardini, A., Girardi, M., \& Gottlober, S., 1997, New Astronomy, 2 119

\bibitem{}Bunn, E.F., \& White, M., 1997, ApJ, 480, 6

\bibitem{}Carlberg, R.G., Morris, S.M., Yee, H.K.C., \& Ellingson, E., 1997a,
ApJ, 479, L19

\bibitem{}Carlberg, R.G., Yee, H.K.C., \& Ellingson, E., 1997b, ApJ, 478, 462

\bibitem{}Carlstrom, J., 1997, Proc. National Academy of Sciences (in press)

\bibitem{}Donahue, M., 1996, ApJ, 468, 79

\bibitem{}Donahue, M., Gioia, I.M., Luppino, G., Hughes, J.P., \& Stocke, J.T.,  1997, astro-ph/9707010

\bibitem{}Edge, A.C., Stewart, G.C., Fabian, A., \& Arnaud, K.A., 1990, MNRAS, 245, 559

\bibitem{}Eke, V.R., Cole, S., \& Frenk, C.S., 1996, MNRAS, 282, 263

\bibitem{}Evrard, A.E., 1989, ApJ, 341, L71

\bibitem{}Evrard, A.E., Metzler, C.A., \& Navarro, J.F., 1996, ApJ, 469, 494

\bibitem{}Fadda, A.D., Girardi, M., Giuricin, G., Mardirossian, F., \& Messetti, M.,  1996, ApJ, 473, 670 

\bibitem{}Fan, X., Bahcall, N.A., \& Cen, R., 1997, ApJ, 490, L123

\bibitem{}Fisher, P., \& Tyson, J.A., 1997, AJ, 114, 14

\bibitem{}Henry, J.P., \& Arnaud, K.A., 1991, ApJ, 372, 410

\bibitem{}Henry, J.P., Gioia, I.M., Maccacaro, T., Morris, S.L., Stocke, J.T.,
 \& Wolter, A., 1992, ApJ, 386, 408

\bibitem{}Henry, J.P., 1997, ApJ, 489, L1

\bibitem{}Hjorth, J., Oukbir, J., \& Kampen, E., 1998, MNRAS, submitted

\bibitem{}Kitayama, T, \& Suto, Y., 1996, ApJ, 469, 480

\bibitem{}Luppino, G.A., \& Gioia, I.M., 1995, ApJ, 445, L77

\bibitem{}Luppino, G.A., \& Kaiser, N.,  1997, ApJ, 475, 20

\bibitem{}Mazure, A., {\em et al.}, 1996, A\&A, 310, 31

\bibitem{}Mushotzky, R.F., \& Scharf, C.A., 1997, ApJ, 482, L13

\bibitem{}Oukbir, J., \& Blanchard, A., 1992, A\&A, 262, L21

\bibitem{}Oukbir, J., \& Blanchard, A., 1997, A\&A, 317, 1

\bibitem{}Peebles, P.J.E., Daly, R.A., \& Juszkiewicz, R., 1989, ApJ, 347, 563

\bibitem{}Peebles, P.J.E., 1993, {\em Principles of Physical Cosmology} (Princeton Univ. Press : Princeton)

\bibitem{}Pen, U.-L., 1998, ApJ, in press (astro-ph/9610147, v3)

\bibitem{}Press, W.H., \& Schechter, P., 1974, ApJ, 187, 425

\bibitem{}Smail, I., Ellis, R.S., Fitchett, M.J., \& Edge, A.C., 1995,
MNRAS, 273, 277

\bibitem{}Viana, P.P., \& Liddle, A.R., 1996, MNRAS, 281, 323

\bibitem{}White, S.D.M., Efstathiou, G., \& Frenk, C.S., 1993, MNRAS, 262, 1023
\end{thebibliography}
\end{document}